\documentclass[cits]{PoS}

\usepackage{bbm}
\usepackage{amsmath, amsthm, amsfonts, amssymb}
\usepackage{mathrsfs}
\usepackage{dsfont}
\usepackage{graphicx}
\usepackage{wrapfig}
\usepackage{minibox}
\usepackage{xcolor}

\newcommand\be            {\begin{equation}}
\newcommand\ee            {\end{equation}}
\newcommand\Renyi         {R\'enyi\ }
\newcommand\Tr            {\mathrm{Tr}}
\newcommand{\mt}[1]       {\textrm{\tiny #1}}
\newcommand\scr           {\scriptstyle}
\newcommand\arxiv[2]      {\href{http://arXiv.org/abs/#1}{#2}}
\newcommand\doi[2]        {\href{http://dx.doi.org/#1}{#2}}

\title{Holographic Calculation of \Renyi Entropies\\and Restrictions on Higher Derivative Terms}

\ShortTitle{Holographic \Renyi Entropies and Restrictions on Higher Derivative Terms}

\author{\speaker{Georgios Pastras}\\
        Department of Physics, School of Applied Mathematics and Physical Sciences,\\National Technical University, Athens 15780, Greece\\
        E-mail: \email{pastras@email.ntua.gr}}
\author{Dimitrios Manolopoulos\\Department of Mathematics, University of Patras, Patras 26110, Greece\\
		Department of Nuclear and Particle Physics, Faculty of Physics,\\University of Athens,
Athens 15784, Greece\\
		E-mail: \email{dmanolop@phys.uoa.gr}}

\abstract{We perform a holographic calculation of the Entanglement \Renyi entropy $S_q(\mu,\lambda)$, for spherical entangling surfaces in boundary CFT's with Einstein-Gauss-Bonnet-Maxwell holographic gravitational duals. We find that for Gauss-Bonnet couplings $\lambda$, larger than a specific value, but still allowed by causality, a violation of an inequality that \Renyi entropies must obey by definition occurs. This violation is related to the existence of negative entropy black holes and restricts the coefficient of the Gauss-Bonnet coupling in the bulk theory. Furthermore, we discover a distinction in the analytic structure of the analytic continuation of $S_q(\mu,\lambda)$, between negative and non-negative $\lambda$, suggesting the existence of a phase transition.}

\FullConference{Proceedings of the Corfu Summer Institute 2014 "School and Workshops on Elementary Particle Physics and Gravity",\\
		3-21 September 2014\\
		Corfu, Greece }

\begin{document}

\section{Introduction}
Quantum entanglement is a physical phenomenon that occurs when composite quantum systems are generated or interact in such a way that the quantum state of each subsystem cannot be described independently, but rather only a quantum state for the system as a whole can be defined. When quantum entanglement is present, measurements of physical quantities in the entangled quantum subsystems are correlated. The most famous such example was described in \cite{Einstein:1935rr}, which later became known as the EPR paradox. Initially, such behaviour was considered contradictory to local causality and thus a reason to consider the quantum description of the world as incomplete, until this counter-intuitive behaviour was verified experimentally in measurements of polarization or spin of entangled particles.

As stated above, an entangled quantum subsystem cannot be described by a wavefunction, but rather a density matrix, which can be calculated from the overall density matrix by tracing out the degrees of freedom of the complement of the subsystem under study. A subsystem which is not entangled with its environment is described by a density matrix corresponding to a pure state, while an entangled subsystem is described by a density matrix corresponding to a mixed state. As a consequence, entanglement is encoded in the spectrum of this density matrix. A naive choice for an entanglement measure is the Shannon entropy applied to the spectrum of the reduced density matrix describing the subsystem under study, known in the literature as the Entanglement Entropy. Entanglement entropy, like Shannon entropy, admits several extensions to a family of entropies like \Renyi entropy \cite{Renyi:1961,Renyi:1965} or Tsallis entropy, providing more information about the probability distribution on which they are defined or convenience in calculations. Entanglement entropy and its \Renyi extension, which from now on we will call Entanglement \Renyi Entropy, have found a series of applications ranging from condensed matter physics \cite{Levin:2006zz,Kitaev:2005dm,Hamma:2005,Calabrese:2004eu,Calabrese:2005zw} to quantum computation \cite{Abramsky:2003} and more recently to quantum gravity and holography \cite{Ryu:2006bv,Ryu:2006ef,Nishioka:2009un,Takayanagi:2012kg,VanRaamsdonk:2009ar,VanRaamsdonk:2010pw,Bianchi:2012ev,Myers:2013lva,Balasubramanian:2013rqa}.

Entanglement entropy in the framework of holography first appeared in \cite{Ryu:2006bv}, where a fascinating conjecture was proposed, named after Ryu and Takayanagi, relating the entanglement entropy of a spacelike region $A$ in the boundary theory with the area of a minimal surface in the bulk with the same boundary as region $A$. This relation presents an intriguing similarity with the area law for black hole entropy, although this minimal surface is not necessarily related with some kind of event horizon. RT conjecture can be seen as a quantitative tool to understand the emergence of gravitational dynamics from quantum statistical physics related to entanglement, explaining the similarity between gravity and thermodynamics. Although the RT conjecture has not been proved yet, there are significant results supporting its validity \cite{Ryu:2006ef,Headrick:2010}.

The calculation of entanglement entropy in field theory is not an easy task, because of the difficulty to express the logarithm of the reduced density matrix. The standard approach which bypasses this difficulty involves the so called Replica Trick and the \Renyi extension of entanglement entropy \cite{Calabrese:2004eu,Calabrese:2005zw}. The RT conjecture has provided more opportunities for the calculation of entanglement entropy in conformal field theories with gravitational holographic duals. An interesting approach \cite{Casini:2011kv}, applicable for regions defined by spherical entangling surfaces, relates the entanglement entropy to the thermal entropy in a hyperbolic cylinder through a conformal transformation. Here, we use this approach as generalized for the grand canonical ensemble \cite{Caputa:2013,Belin:2013uta}, following closely the work in \cite{Pastras:2014oka}.

\section{Entanglement \Renyi Entropy}
\label{sec:renyi}

\subsection{Entropy and Entanglement}
\label{subsec:entanglement}

\begin{figure}[ht!]
\[
\raisebox{-58pt}{
  \begin{picture}(50,140)
    \put(-90,3){\scalebox{4}{\includegraphics{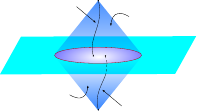}}}
  \put(-49,7){
    \setlength{\unitlength}{.90pt}\put(-32,-1){
    \put(153,115) {$\scr \mathcal{D}^+(A)$}
    \put(45,15) {$\scr \mathcal{D}^-(A)$}
    \put(110,143) {$\scr p$}
    \put(98,65) {$\scr q$}
    \put(125,65) {$\scr q'$}
    \put(145,65) {$\scr A$}
    \put(50,57) {$\scr \partial A$}
    \put(110,-10) {$\scr p'$}
    \put(202,47) {$\scr A^c$}
    \put(-10,35) {$\scr t=0$ \scriptsize{hypersurface}}
    \put(150,0) {\scriptsize{\minibox{future directed \\ timelike curve \\ from $p'$}}}
    \put(20,130) {\scriptsize{\minibox{past directed \\ timelike curve \\ from $p$}}}
         }\setlength{\unitlength}{1pt}}
  \end{picture}}
\]
\caption{\small{The division of the $t=0$ time slice into two regions $A$ and $A^c$ by the entangling surface $\partial A$. The figure also depicts the Cauchy development $\mathcal{D}(A)=\mathcal{D}^+(A)\cup \mathcal{D}^-(A)$.}}\label{fig:Cauchy development}
\end{figure}
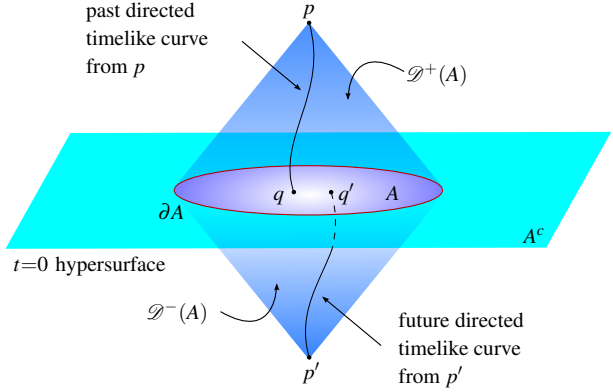

We consider a quantum system in $d$-dimensional Minkowski spacetime $\mathds{R}^{1,d-1}$. This system is divided to two subsystems $A$ and $A^c$ by a closed codimension 2 surface $\partial A$ at time $t=0$, as in figure \ref{fig:Cauchy development}. Then, system $A$ is described by the density matrix $\rho_\mt{A}$, which is derived from the overall density matrix $\rho$ by tracing out the degrees of freedom of $A^c$.
\begin{equation}
\rho_\mt{A}  = \Tr_{\mt{A}^c} \rho .
\end{equation}

For later use, we also note that the Cauchy development $\mathcal{D}(A)$ is defined as the set of all space-time events $p\in\mathds{R}^{1,d-1}$ with the property that every non-spacelike curve through $p$ intersects $A$ at least once (figure \ref{fig:Cauchy development}).

When entanglement is present, the density matrix of subsystem $A$ may correspond to a mixed state, even if the overall system lies in a pure state. Let's implement a simple example to clarify how entanglement is encoded into  the spectrum of the density matrix $\rho_\mt{A}$.

Consider a composite quantum system being composed of just two spinors. If the whole system lies in a non-entangled pure state, for example, $\left|\psi\right\rangle = \frac{1}{2}\left({\left({\left|\uparrow\right\rangle + \left|\downarrow\right\rangle}\right)_\mt{A}}\otimes{\left({\left|\uparrow\right\rangle + \left|\downarrow\right\rangle}\right)_{\mt{A}^c}}\right)$, then the reduced density matrix $\rho_\mt{A}$ of subsystem $A$ is given by $\rho_\mt{A} = \frac{1}{2}\left({\left|\uparrow\right\rangle + \left|\downarrow\right\rangle }\right)\left({\left\langle\uparrow\right| + \left\langle\downarrow\right|}\right)$, which has eigenvalues equal to 0 and 1 and thus, describes a pure state. On the contrary, if the whole system lies in the maximally entangled pure state $\left| \psi  \right\rangle  = \frac{1}{{\sqrt 2 }}\left( {{{\left|  \uparrow  \right\rangle }_\mt{A}} \otimes {{\left|  \uparrow  \right\rangle }_{\mt{A}^c}} + {{\left|  \downarrow  \right\rangle }_\mt{A}} \otimes {{\left|  \downarrow  \right\rangle }_{\mt{A}^c}}} \right)$, then the density matrix $\rho_\mt{A}$ of subsystem $A$ is given by ${\rho _\mt{A}} = \frac{1}{2}\left( {\left|  \uparrow  \right\rangle \left\langle  \uparrow  \right| + \left|  \downarrow  \right\rangle \left\langle  \downarrow  \right|} \right)$, which has two eigenvalues equal to $\frac{1}{2}$ and thus, it corresponds to a mixed state.

The example strongly suggests that the entanglement between subsystems $A$ and $A^c$ is encoded into the spectrum of $\rho_\mt{A}$ and specifically is correlated with its ``disorder''. Following the paradigm of statistical mechanics, we adopt a simple logarithmic measure of this ``disorder'', namely the Shannon entropy, and we define the entanglement entropy as,
\begin{equation}
S_\mt{EE}  :=  - {\mathop{\rm Tr}\nolimits} \left( {\rho_\mt{A} \ln \rho _\mt{A} } \right) .
\label{eq:entanglement_entropy_definition}
\end{equation}

We have to notice that this is a useful definition under a hidden implicit assumption. Specifically, the overall system has to lie in a pure state. For example, one could take the same reduced density matrix $\rho_\mt{A}$ as in the case of the maximally entangled overall state, had they considered the overall system lying in the mixed state, $\rho  = \frac{1}{2}\left( {{{\left|  \uparrow  \right\rangle }_\mt{A}} \otimes {{\left|  \uparrow  \right\rangle }_{\mt{A}^c}}{{\left\langle  \uparrow  \right|}_\mt{A}} \otimes {{\left\langle  \uparrow  \right|}_{\mt{A}^c}} + {{\left|  \downarrow  \right\rangle }_\mt{A}} \otimes {{\left|  \downarrow  \right\rangle }_{\mt{A}^c}}{{\left\langle  \downarrow  \right|}_\mt{A}} \otimes {{\left\langle  \downarrow  \right|}_{\mt{A}^c}}} \right)$. Thus, when physical configurations where the overall system lies in a mixed state, such as a thermal state, are considered, one has to be very careful about whether entanglement entropy really measures entanglement.

\subsection{Entanglement \Renyi Entropy}
\label{subsec:renyi_basics}

A technical problem that usually appears in various calculations of entanglement entropy is the difficulty of representing the operator $\ln \rho_\mt{A}$, which appears in the definition of $S_\mt{EE}$. On the contrary, if simple states for the overall system are considered, such as the ground state or a thermal ensemble, the reduced density matrix $\rho_\mt{A}$ and its positive integer powers can be expressed as path integrals. It is thus a good idea to implement the \Renyi extension of Shannon entropy to define the entanglement \Renyi entropy as
\begin{equation}
S_q := \frac{1}{{1 - q}}\ln {\mathop{\rm Tr}\nolimits} \rho_\mt{A} ^q , \ \ q\in\mathds{R}^+ -\{1\}
\label{eq:Renyi_entropy_definition}
\end{equation}
and then recover the entanglement entropy in the limit
\begin{equation}
S_\mt{EE} = \mathop {\lim }\limits_{q \to 1} S_q .
\end{equation}

Another advantage of the introduction of entanglement \Renyi entropies is the fact that they contain much more information about the spectrum of $\rho_\mt{A}$ than entanglement entropy alone. Specifically the knowledge of $S_q$ for all $q>0$ is sufficient to recover the whole spectrum of $\rho_\mt{A}$ \cite{Calabrese:2008es}.

\Renyi entropies obey the following four inequalities by definition
\begin{equation}
\frac{{\partial {S_q}}}{{\partial q}} \le 0,\quad
\frac{\partial }{{\partial q}}\left( {\frac{{q - 1}}{q}{S_q}} \right) \ge 0, \quad
\frac{\partial }{{\partial q}}\left( {\left( {q - 1} \right){S_q}} \right) \ge 0,\quad
\frac{{{\partial ^2}}}{{\partial {q^2}}}\left( {\left( {q - 1} \right){S_q}} \right) \le 0.
\label{eq:ren inequalities}
\end{equation}
Later on, we will perform a holographic calculation of \Renyi entropies in CFTs from holographic dual theories. As such calculations are performed indirectly from quantities related to the dual bulk theory, the validity of these inequalities is not obvious and can provide interesting constraints.

\subsection{\Renyi Entropy in the Grand Canonical Ensemble}
\label{subsec:renyi_grand_canonical}

We would like to consider theories with a conserved charge in the grand canonical ensemble. Thus, an extension of entanglement \Renyi entropy is required, as described in \cite{Caputa:2013,Belin:2013uta},
\begin{equation}\label{eq:charged_Renyi_entropy_definition}
S_q (\mu) := \frac{1}{{1 - q}}\ln \Tr\varrho_\mt{A}^q(\mu) , \ \ \ \varrho_\mt{A}(\mu)\equiv{\rho _\mt{A} \frac{{e^{\mu Q_\mt{A} } }}{{n_\mt{A} \left( \mu  \right)}}},
\end{equation}
where $n_\mt{A}(\mu)\equiv \Tr\left[ {\rho _\mt{A} e^{\mu Q_\mt{A} } } \right]$. It is also interesting to define the analytical continuation of these entanglement \Renyi entropies for an imaginary chemical potential
\begin{equation}\label{eq:charged_Renyi_entropy_definition_imaginary}
\tilde{S}_q (\mu_\mt{E}) := \frac{1}{{1 - q}}\ln \Tr\tilde{\varrho}_\mt{A}^q(\mu_\mt{E}) , \ \ \ \tilde{\varrho}_\mt{A}(\mu_\mt{E})\equiv\varrho_\mt{A}(i\mu_\mt{E}) ,
\end{equation}
with $\mu_\mt{E}\in\mathds{R}$. When we analytically continue from \eqref{eq:charged_Renyi_entropy_definition} to \eqref{eq:charged_Renyi_entropy_definition_imaginary}, typically, only some singularities appear in the imaginary $\mu$-axis.

\section{Holographic Computation of \Renyi Entropy}
\label{sec:holographic}

In this section, following \cite{Casini:2011kv,Belin:2013uta}, we review some aspects of charged \Renyi entropies for a quantum system in $d$-dimensional Euclidean space-time $\mathds{R}^d$. This quantum system consists of two subsystems $A$ and $A^c$, which are separated by a spherical entangling surface $S^{d-2}$.

In \cite{Casini:2011kv}, it was shown that the entanglement entropy is related to the thermal entropy $S(T)$ of the CFT on the hyperbolic cylinder $\mathds{R}\times\mathds{H}^{d-1}$ with the temperature fixed by the radius of curvature $R$ of $S^{d-2}$. This is shown by conformally mapping the Cauchy development $\mathcal{D}(A)$ of the region $A$ on the inside of $S^{d-2}$ to $\mathds{R}\times\mathds{H}^{d-1}$. In this way, the vacuum correlators in $\mathcal{D}(A)$ are conformally mapped to thermal correlators in $\mathds{R}\times\mathds{H}^{d-1}$. Under this conformal transformation, the reduced density matrix $\rho_\mt{A}$ is mapped to a thermal bath in $\mathds{R}\times\mathds{H}^{d-1}$.

\subsection{Charged \Renyi Entropy for Spherical Entangling Surfaces}
\label{subsec:Charged Ren Entropy}

We start by performing the conformal transformation which maps $\mathcal{D}(A)$ to the Euclidean space $S^1\times\mathds{H}^{d-1}$, where the time coordinate $t_\mt{E}$ is compactified on $S^1$. Consider the flat space metric in polar coordinates:
\be\label{eq:metric R^d in polar form}
ds^2_{\mathds{R}^d}= dt_\mt{E}^2 + dr^2 + r^2dS_{d-2}^2,
\ee
where $r$ is the radial coordinate and $dS_{d-2}^2$ is the metric on $S^{d-2}$. The spherical entangling surface is described by $(t_\mt{E},r)=(0,R)$. One can rewrite this in a complex form by using a complex coordinate $z = r+it_\mt{E}$ and then by performing the coordinate transformation $e^{-w}={(R-z)}/{(R+z)}$, where $w = u + i\frac{\tau_\mt{E}}{R}$, the metric becomes
\be\label{eq:metric R^d-1 to H^d-1}
ds^2_{S^1\times\mathds{H}^{d-1}}= \Omega^2 ds^2_{\mathds{R}^d} = d\tau_\mt{E}^2 + R^2\left(du^2 + \sinh^2u \ dS_{d-2}^2\right),
\ee
where $\Omega = {2R^2}/{\left| R^2 - z^2 \right|} = \left| 1 + \cosh w \right|$ is the conformal factor. Hence, we have a conformal mapping from $\mathcal{D}(A)$ to $S^1\times\mathds{H}^{d-1}$. In particular, it maps the CFT in the Minkowski vacuum on $\mathcal{D}(A)$ to a thermal state on $\mathds{R}\times\mathds{H}^{d-1}$ with temperature $\beta_0^{-1} \equiv T_0 = {1}/{(2 \pi R)}$.

The thermal density matrix in the new spacetime $S^1\times\mathds{H}^{d-1}$ is related to the reduced density matrix on $\mathcal{D}(A)$ by a unitary transformation $U$ as,
\be\label{eq:unitary transform or rho}
\rho_\mt{A} \mapsto U^{-1}\frac{e^{-\beta_0 H}}{Z(\beta_0)}U, \ \ \mathrm{where} \ \ Z(\beta) \equiv \Tr \ e^{-\beta H}.
\ee
Thus, the entanglement entropy across the sphere is equal to the thermal entropy $S(T)$ in $\mathds{R}\times\mathds{H}^{d-1}$. 

When considering charged \Renyi entropies as in \cite{Caputa:2013,Belin:2013uta}, the current operator $J_\mu$ in the underlying CFT has to be considered. This operator is associated with the conserved global charge
\be\label{eq:global charge}
Q_\mt{A} = \int_\mt{A} \! d^{d-1}x \ J_0.
\ee
This global symmetry in the boundary CFT, according to \cite{Belin:2013uta}, gives rise to a gauge field in the dual gravitational theory and thus, $S_q(\mu)$ is related to the thermal entropy of a charged topological black hole. In terms of the boundary CFT, this means that in order to extend the path integral calculations of $S_q$ to include a chemical potential, one needs to insert a Wilson loop encircling the entangling surface. The relevant fixed background gauge potential $B_\mu$, coupled to the conserved current $J_\mu$, represents the chemical potential. However, in the thermal path integral, the Euclidean time direction is compactified with period $\beta_0 \equiv T^{-1}_0$ and then the chemical potential appears by inserting a nontrivial Wilson line
\be\label{eq:wilson loop exp}
W(\mathcal{C}):= \exp \left(i Q_\mt{A}\oint_{\mathcal{C}}\! dx^\mu \ B_\mu \right) = e^{i\mu_\mt{E} Q_\mt{A}},
\ee
on this thermal circle. Note, that since we are discussing the abelian case, we do not need the trace and the exponential does not need to be path ordered.

The conformal mapping which leads to the definition of the charged \Renyi entropy in a \emph{grand canonical ensemble} with an imaginary chemical potential is
\be\label{eq:unitary transform or rho grand canonical}
\tilde{\varrho}_\mt{A}^q(\mu_\mt{E}) \mapsto U^{-1}\tilde{\varrho}^q_\mt{A}(\beta_0,\mu_\mt{E})U =  U^{-1}\frac{e^{-q\left(\beta_0 H-i\mu_\mt{E} Q_\mt{A}\right)}}{\tilde{Z}^q(\beta_0,\mu_\mt{E})}U, \ \ \ \tilde{Z}(\beta,\mu_\mt{E}) \equiv \Tr \left[e^{-\beta \left(H-i\frac{\mu_\mt{E}}{\beta_0} Q_\mt{A}\right)}\right],
\ee
where $\tilde{Z}(\beta,\mu_\mt{E})$ is the usual grand partition function. As before, $U$ is a unitary transformation, which, upon taking the trace of the above expression, cancels with its inverse to give
\be\label{eq:Tr unitary transform or rho grand canonical}
\Tr \tilde{\varrho}^q_\mt{A}(\mu_\mt{E}) = \frac{\tilde{Z}(q\beta_0,\mu_\mt{E})}{\tilde{Z}^q(\beta_0,\mu_\mt{E})}.
\ee
Therefore, equation \eqref{eq:charged_Renyi_entropy_definition_imaginary} yields
\be\label{eq:charged_Renyi_entropy d-cft}
\tilde{S}_q(\mu_\mt{E}) = \frac{1}{q-1}\left( q \ln\tilde{Z}(\beta_0,\mu_\mt{E}) - \ln\tilde{Z}(q\beta_0,\mu_\mt{E})\right).
\ee

The thermal entropy $\tilde{S}(T,\mu_\mt{E})$ in the grand canonical ensemble is given in terms of the temperature derivative of the free energy
\be\label{eq:S(T,mu)}
\tilde{S}(T,\mu_\mt{E}) = -\left(\frac{\partial \tilde{F}(T,\mu_\mt{E})}{\partial T}\right)_{\mu_\mt{E}} = \left(\frac{\partial}{\partial T}\left( T\ln\tilde{Z}(T^{-1},\mu_\mt{E}) \right)\right)_{\mu_\mt{E}}.
\ee
One therefore arrives at the following relation between charged \Renyi entropy and thermal entropy
\begin{equation}
\tilde{S}_q \left( {\mu_\mt{E}} \right) = \frac{q}{{q - 1}}\frac{1}{{T_0 }}\int_{T_0/q}^{T_0 } \! {dT \ \tilde{S} \left( {T,\mu_\mt{E} } \right)} .
\label{eq:Renyi_replica_trick}
\end{equation}
The above analysis is similar in the case of a real chemical potential.

In this study, we investigate CFTs whose holographic dual is Einstein-Gauss-Bonnet gravity. The introduction of the Gauss-Bonnet coupling allows the investigation of a broader class of CFTs with unequal central charges. Standard AdS/CFT dictionary relates the thermal entropy in the hyperbolic cylinder to the entropy of asymptotically AdS black holes with hyperbolic horizons. The latter can be easily calculated using standard black hole thermodynamics tools \cite{Wald:1993nt,Jacobson:1993vj,Iyer:1994ys}. Moreover, the AdS/CFT dictionary upgrades the global symmetry of the boundary theory to a gauge symmetry in the bulk. Thus, thermodynamics of asymptotically AdS charged black holes with hyperbolic horizons in Einstein-Gauss-Bonnet gravity, already known in the literature \cite{Cvetic:2001bk,Ge:2008ni,Anninos:2008sj}, combined with \eqref{eq:Renyi_replica_trick} allow the calculation of entanglement \Renyi entropies based solely on quantities of the bulk theory.

It is more convenient to express all the thermodynamic quantities as function of the horizon radius normalized by the AdS curvature scale, $x \equiv {{r_\mt{H} }}/{{\tilde L}} $. In this case, formula \eqref{eq:Renyi_replica_trick}, which yields entanglement \Renyi entropy is be expressed as
\begin{equation}
S_q \left( {\mu,\lambda } \right) = \frac{q}{{q - 1}}\frac{1}{{T_0 }}\int_{x_q }^{x_1 } \! {dx \ S\left( {x;\lambda } \right) {\partial_x T\left( {x;\mu,\lambda } \right)}},  \\
\label{eq:Renyi_entropy_formula}
\end{equation}
where $\lambda$ is the Gauss-Bonnet coupling and $x_q$ is the largest real solution of $T\left( {x_q ,\mu } \right) = {{T_0 }}/{q}$.

\subsection{Einstein-Gauss-Bonnet-Maxwell Gravity}
\label{subsec:EGBM}

Einstein-Gauss-Bonnet gravity coupled with electromagnetism is described by the action,
\begin{equation}
\label{eq:einstein_gauss_bonnet_maxwell_action}
 I = \frac{1}{{2\ell_P ^{d - 1} }}\int \! {d^{d + 1} x} \ \sqrt { - g} \left( {\frac{{d\left( {d - 1} \right)}}{{L^2 }} + R}
 { + \frac{{\lambda L^2 }}{{\left( {d - 2} \right)\left( {d - 3} \right)}}\mathcal{X}- \frac{{\ell_* ^2 }}{4}F_{\mu \nu } F^{\mu \nu } } \right),
\end{equation}
where $\mathcal{X} \equiv R^2  - 4R_{\mu \nu } R^{\mu \nu }  + R_{\mu \nu \kappa \lambda } R^{\mu \nu \kappa \lambda }$  is the Gauss-Bonnet term.

An interesting feature of this class of theories is that the central charges of the corresponding CFT are not equal. Specifically, in four dimensions \cite{Myers:2010jv},
\begin{equation}
c = {\pi ^2}{\left( {\frac{{\tilde L}}{{{\ell_P}}}} \right)^3}\left( {1 - 2\lambda {f_\infty }} \right) ,\quad
a = {\pi ^2}{\left( {\frac{{\tilde L}}{{{\ell_P}}}} \right)^3}\left( {1 - 6\lambda {f_\infty }} \right) .
\end{equation}
$\tilde L$ denotes the curvature length scale of AdS space, while $f_\infty$ is defined as $f_\infty = L^2 / {\tilde L}^2$.
In higher dimensions, following \cite{Hung:2011nu}, we use the central charge given by
the leading singularity of the two-point function of the stress tensor \cite{Buchel:2009sk} and the central charge defined in \cite{Myers:2010tj,Myers:2010xs}, which obeys a holographic c-theorem
\begin{equation}\label{eq:central charge of T}
{{\tilde C}_T} = \frac{{{\pi ^{\frac{d}{2}}}}}{{\Gamma \left( {\frac{d}{2}} \right)}}{\left( {\frac{{\tilde L}}{{{\ell_P}}}} \right)^{d - 1}}\left( {1 - 2\lambda {f_\infty }} \right), \quad
{a_d}^* = \frac{{{\pi ^{\frac{d}{2}}}}}{{\Gamma \left( {\frac{d}{2}} \right)}}{\left( {\frac{{\tilde L}}{{{\ell_P}}}} \right)^{d - 1}}\left( {1 - 2\frac{{d - 1}}{{d - 3}}\lambda {f_\infty }} \right) .
\end{equation}

Gravity with Gauss-Bonnet corrections may contain negative energy excitations, leading to causality violation \cite{Hofman:2008ar}. Demanding causality constrains the range of the $\lambda$ \cite{Buchel:2009sk,Camanho:2009},
\begin{equation}
 - \frac{{\left( {3d + 2} \right)\left( {d - 2} \right)}}{{4{{\left( {d + 2} \right)}^2}}} \le \lambda  \le \frac{{\left( {d - 2} \right)\left( {d - 3} \right)\left( {{d^2} - d + 6} \right)}}{{4{{\left( {{d^2} - 3d + 6} \right)}^2}}} , \quad \textrm{or} \quad \frac{{d\left( {d - 3} \right)}}{{{d^2} - 2d - 2}} \le \frac{{{{\tilde C}_T}}}{{{a_d}^*}} \le \frac{d}{2}
\label{eq:lambda_causality_bound}
\end{equation}
in terms of the central charges of the boundary CFT.

The holographic dual of the grand canonical ensemble in the boundary CFT are charged topological black holes with hyperbolic horizons. The Hawking temperature of these solutions can be expressed in terms of the horizon radius and the chemical potential \cite{Cvetic:2001bk,Ge:2008ni,Anninos:2008sj},
\begin{equation}
T\left( {x;\mu,\lambda} \right) = \frac{T_0}{2{f_\infty }}\frac{1}{x}\frac{{d{x^4} - \left( {d - 2} \right)\left( {1 + \frac{{d - 2}}{{2\left( {d - 1} \right)}}{{\left( {\frac{{\mu {\ell_*}}}{{2\pi \tilde{L}}}} \right)}^2}} \right){f_\infty }{x^2} + \left( {d - 4} \right)\lambda {f_\infty }^2}}{{{x^2} - 2\lambda {f_\infty }}} .
\label{eq:black_hole_temperature}
\end{equation}
The black hole entropy can be calculated using Wald's formula \cite{Wald:1993nt,Jacobson:1993vj,Iyer:1994ys}  and is found equal to
\begin{equation}
{S}\left( {x;\lambda} \right) = {V_{\mathds{H}^{d-1}} }{\left( {\frac{{\tilde L}}{{{\ell_P}}}} \right)^{d - 1}}2\pi \left( {x^{d - 1} - 2\frac{{d - 1}}{{d - 3}}\lambda {f_\infty }x^{d - 3}} \right) .
\label{eq:black_hole_thermal_entropy}
\end{equation}
Finally, equation \eqref{eq:black_hole_temperature} implies that the quantity $x_q$, which satisfies $T\left( {x_q ,\mu } \right) = {{T_0 }}/{q}$, is the largest real solution of equation
\begin{equation}
d\frac{{x_q^4}}{{{f_\infty }}} - 2\frac{{x_q^3}}{q} - \left( {d - 2} \right)\left( {1 + \frac{{d - 2}}{{2\left( {d - 1} \right)}}{{\left( {\frac{{\mu {\ell_*}}}{{2\pi \tilde{L}}}} \right)}^2}} \right)x_q^2 + 4\lambda {f_\infty }\frac{{x_q}}{q} + \left( {d - 4} \right)\lambda {f_\infty } = 0 .
\label{eq:xn_equation}
\end{equation}
It is easy to recover the results of \cite{Hung:2011nu} in the vanishing chemical potential limit and the results of \cite{Belin:2013uta} in the vanishing Gauss-Bonnet coupling limit.

\subsection{Holographic Charged \Renyi Entropies}
\label{subsec:Renyi}

It is straightforward to calculate the \Renyi entropy on the basis of equation \eqref{eq:Renyi_entropy_formula}, using equations \eqref{eq:black_hole_temperature} and \eqref{eq:black_hole_thermal_entropy}. Equation \eqref{eq:Renyi_replica_trick} suggests that at the limit $q\to 1$, corresponding to the entanglement entropy, one should recover the entropy of the topological black hole with temperature $T_0$. This is consistent with the Ryu-Takayanagi conjecture for entanglement entropy \cite{Ryu:2006bv,Ryu:2006ef}, which claims that the entanglement entropy for a region $A$ equals
\begin{equation}
{S_1}\left( A \right) = \frac{{{\rm{Area}}\left( {{m_\mt{A}}} \right)}}{{4G_N}},
\label{eq:RTformula}
\end{equation}
where $m_\mt{A}$ is the minimal open surface in the bulk with boundary $\partial A$. Quantum corrections to the above formula affect the value of the gravitational constant, while $a'$ corrections result in the correction of the area functional, with a different one, matching Wald's formula for black hole entropy \cite{Wald:1993nt,Jacobson:1993vj,Iyer:1994ys}. The conformal transformation used in subsection \ref{subsec:Charged Ren Entropy}, if appropriately extended in the bulk, maps the minimal surface $m_\mt{A}$ to the hyperbolic horizon of the topological black hole. Thus, consistency with Ryu-Takayanagi conjecture requires that the entanglement entropy limit coincides with the entropy of the topological black hole.

Calculation of entanglement \Renyi entropy using formula \eqref{eq:Renyi_entropy_formula} usually involves a by parts integration that may yield a result that does not manifestly have the appropriate $q\to 1$ limit, as in the results of \cite{Belin:2013uta}. However, the result can always be written in a form manifestly consistent with Ryu-Takayanagi conjecture, making use of formula \eqref{eq:xn_equation} to eliminate the explicit dependence of the entanglement \Renyi entropy on the chemical potential,
\begin{multline}
{S_q}\left( {\mu ,\lambda } \right) = V_{\mathds{H}^{d-1}}{\left( {\frac{{\tilde L}}{{{\ell_P}}}} \right)^{d - 1}}2\pi \frac{q}{q - 1}\frac{{d - 1}}{{d - 2}} \left[ {\frac{{x_1^d - x_q^d}}{{{f_\infty }}}} \right. - \lambda {f_\infty }\left( {x_1^{d - 4} - x_q^{d - 4}} \right)\\
\left. { - \frac{1}{{d - 1}}\left( {x_1^{d - 1} - \frac{{x_q^{d - 1}}}{q}} \right) - \frac{{2\lambda {f_\infty }}}{{d - 3}}\left( {x_1^{d - 3} - \frac{{x_q^{d - 3}}}{q}} \right)} \right].
\label{eq:holographic_Renyi_entropy_result_simplified}
\end{multline}
This still depends implicitly on $\mu$ through $x_1$ and $x_q$.

\begin{figure}[ht!]
\[
\raisebox{-58pt}{
  \begin{picture}(0,130)
  \put(-215,0){\scalebox{1.5}{\includegraphics{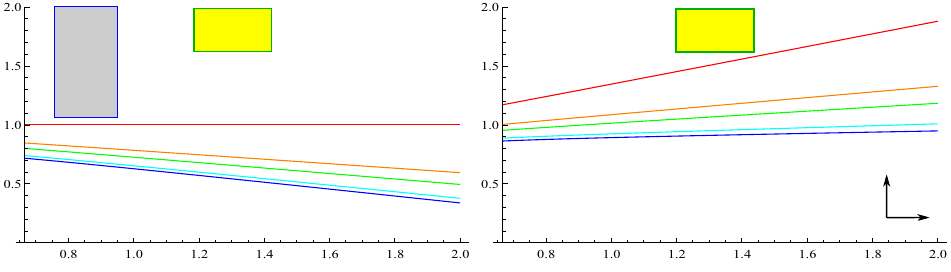}}}
  \put(0,5){
    \setlength{\unitlength}{.90pt}\put(-32,-1){
    \put(196,47) {$\scr S_q(\mu)/S_1(0)$}
    \put(244,21) {$\scr \frac{\tilde{C}_T}{a_d^*}$}
    \put(-178,113) {$\scr \color{red}q=1$}
    \put(-178,103) {$\scr \color{orange}q=2$}
    \put(-178,93) {$\scr \color{green}q=3$}
    \put(-178,83) {$\scr \color{cyan}q=10$}
    \put(-178,73) {$\scr \color{blue}q=100$}
    \put(-109,109) {$\scr \frac{\mu\ell_*}{2\pi \tilde{L}}=0$}
    \put(123,109) {$\scr \frac{\mu\ell_*}{2\pi \tilde{L}}=1$}
      }\setlength{\unitlength}{1pt}}
  \end{picture}}
\]
\caption{\small{\Renyi entropies normalized by $S_1 \left( 0 \right)$ as function of the Gauss-Bonnet coupling for various chemical potentials and $d=4$}.}\label{Fig:3}
\end{figure}
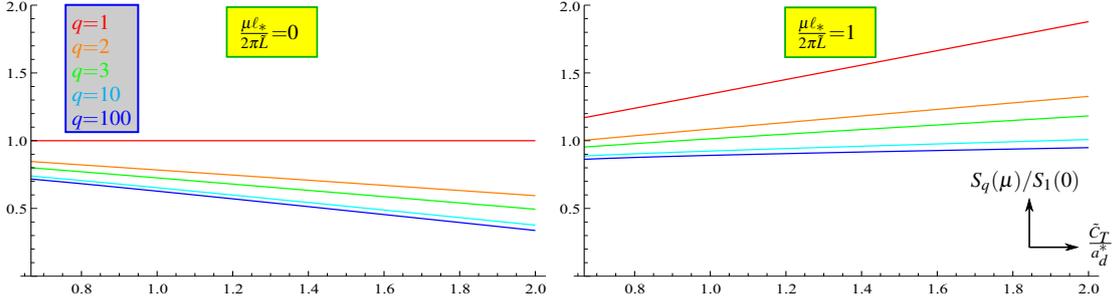

Since the derived expression \eqref{eq:holographic_Renyi_entropy_result_simplified} for entanglement \Renyi entropy is a function of the solution of equation \eqref{eq:xn_equation}, a numerical study of \Renyi entropy can shed more light to its dependence on the central charges and chemical potential. In figure \ref{Fig:3}, the entanglement \Renyi entropy is plotted versus the central charges ratio ${\tilde{C}_T}/{a_d^*}$ covering all the range of the former allowed by causality, as described by \eqref{eq:lambda_causality_bound}. In order to remove the dependence of entanglement \Renyi entropy on the divergent volume $V_{\mathds{H}^{d-1}}$, we regulate it calculating its ratio with the entanglement entropy.
In agreement with \cite{Hung:2011nu}, the dependence of $S_q(\mu)/S_1(0)$ on ${\tilde{C}_T}/{a_d^*}$ is approximately linear. However a significant difference is present; the maximum value of ${S_q \left( \mu \right)}/{S_1 \left( 0 \right)}$ is not always achieved for the minimum allowed value of the ratio ${\tilde{C}_T}/{a_d^*}$. As the chemical potential increases, the ratio ${S_q \left( \mu \right)}/{S_1 \left( 0 \right)}$ for $q$ smaller than a critical value, acquires its maximum value for the maximum allowed value of the ratio ${\tilde{C}_T}/{a_d^*}$, until a critical finite chemical potential, where the above becomes true for all $q$'s. These critical quantities can be estimated by performing a linear expansion of entanglement \Renyi entropy around the Einstein-Maxwell result. More details are provided in \cite{Pastras:2014oka}.

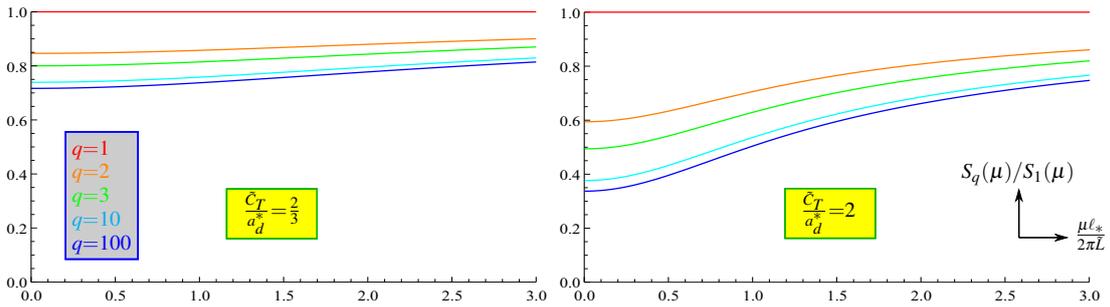
\begin{figure}[h]
\[
\raisebox{-58pt}{
  \begin{picture}(0,130)
  \put(-215,0){\scalebox{1.5}{\includegraphics{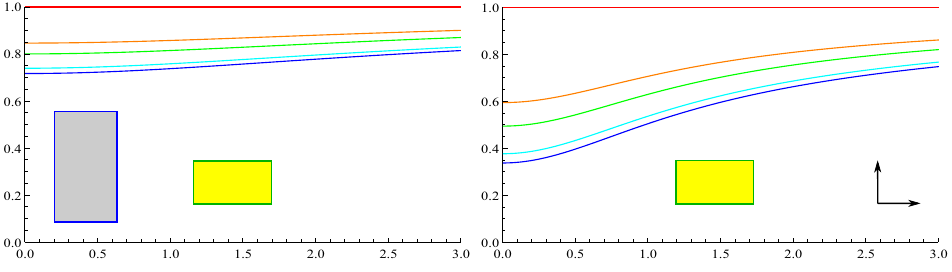}}}
  \put(0,5){
    \setlength{\unitlength}{.90pt}\put(-32,-1){
    \put(192,53) {$\scr S_q(\mu)/S_1(\mu)$}
    \put(239,27) {$\scr \frac{\mu\ell_*}{2\pi \tilde{L}}$}
    \put(-178,63) {$\scr \color{red}q=1$}
    \put(-178,53) {$\scr \color{orange}q=2$}
    \put(-178,43) {$\scr \color{green}q=3$}
    \put(-178,33) {$\scr \color{cyan}q=10$}
    \put(-178,23) {$\scr \color{blue}q=100$}
        \put(-107,37) {$\scr \frac{\tilde{C}_T}{a_d^*}=\frac{2}{3}$}
    \put(125,37) {$\scr \frac{\tilde{C}_T}{a_d^*}=2$}
      }\setlength{\unitlength}{1pt}}
  \end{picture}}
\]
\caption{\small{\Renyi entropies normalized by $S_1 \left( \mu \right)$ as a function of the chemical potential for the minimum and maximum values of ${\tilde{C}_T}/{a_d^*}$ allowed by causality and $d=4$.}}\label{Fig:9}
\end{figure}

An interesting feature of entanglement \Renyi entropies is their asymptotic behaviour for large chemical potentials. It is not difficult to find the asymptotic behaviour of $x_q$ from equation \eqref{eq:xn_equation} to substitute in \eqref{eq:holographic_Renyi_entropy_result_simplified} and find the large chemical potential limit of entanglement \Renyi entropies,
\begin{equation}
\mathop {\lim }\limits_{\mu  \to \infty } {S_q}\left( {\mu ,\lambda} \right) \sim {V_{\mathds{H}^{d-1}}}2\pi{\left( {\frac{{\left( d - 2 \right) \sqrt{f_\infty}}}{{\sqrt {2d\left( {d - 1} \right)} }}\frac{{\mu {\ell _*}}}{{2\pi {\ell _P}}}} \right)^{d - 1}} ,
\end{equation}
which depends only on the product $\sqrt{f_\infty} \mu$ and not on $q$. The latter is evident in figure \ref{Fig:9}.

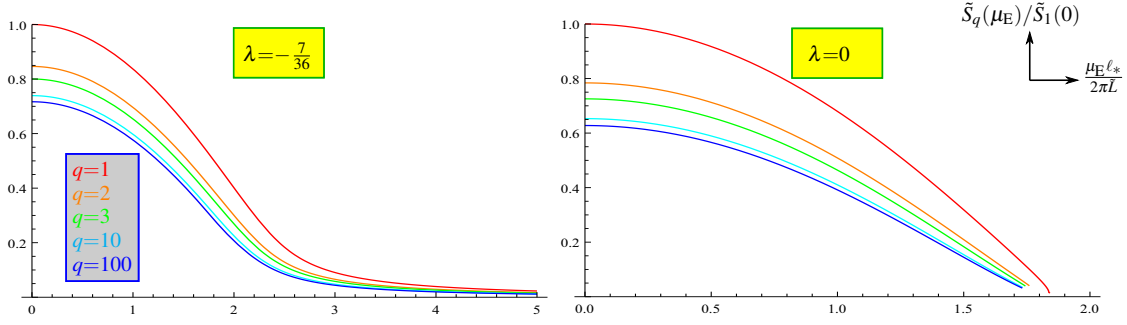
\begin{figure}[h]
\[
\raisebox{-58pt}{
  \begin{picture}(0,130)
  \put(-215,0){\scalebox{1.5}{\includegraphics{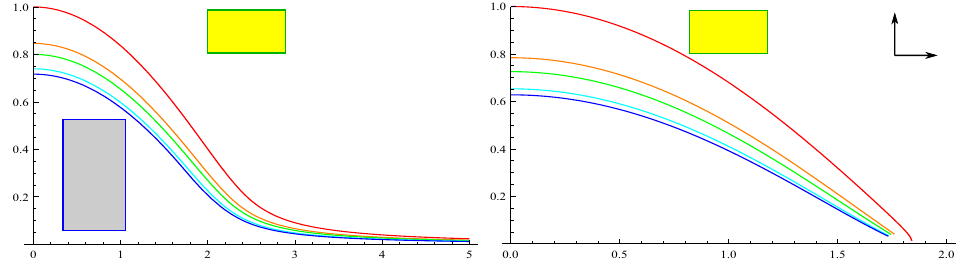}}}
  \put(0,5){
    \setlength{\unitlength}{.90pt}\put(-32,-1){
    \put(196,125) {$\scr \tilde{S}_q(\mu_\mt{E})/\tilde{S}_1(0)$}
    \put(246,99) {$\scr \frac{\mu_\mt{E}\ell_*}{2\pi \tilde{L}}$}
    \put(-174,60) {$\scr \color{red}q=1$}
    \put(-174,50) {$\scr \color{orange}q=2$}
    \put(-174,40) {$\scr \color{green}q=3$}
    \put(-174,30) {$\scr \color{cyan}q=10$}
    \put(-174,20) {$\scr \color{blue}q=100$}
    \put(-103,108) {$\scr \lambda=-\frac{7}{36}$}
    \put(132,108) {$\scr \lambda=0$}
      }\setlength{\unitlength}{1pt}}
  \end{picture}}
\]
\caption{\small{\Renyi entropies normalized by $\tilde{S}_1 \left( 0 \right)$ as a function of the imaginary chemical potential for negative and non-negative Gauss-Bonnet couplings and $d=4$.}}\label{Fig:11}
\end{figure}

The study of the dependence of entanglement \Renyi entropy on an imaginary chemical potential reveals a qualitative difference between non-negative and negative Gauss-Bonnet couplings. As it is visible in figure \ref{Fig:11}, for negative Gauss-Bonnet couplings, entanglement \Renyi entropies are well defined for arbitrarily large imaginary chemical potential and the ratio ${{{{\tilde S}_q}\left( {{\mu_\mt{E}} } \right)}}/{{{{\tilde S}_1}\left( {{0} } \right)}}$ goes to zero for $\mu_\mt{E}$ going to infinity. On the contrary, for non-negative Gauss-Bonnet coupling there is a finite imaginary chemical potential, where equation \eqref{eq:xn_equation} fails to have a real, positive solution implying that the topological black hole becomes a naked singularity.

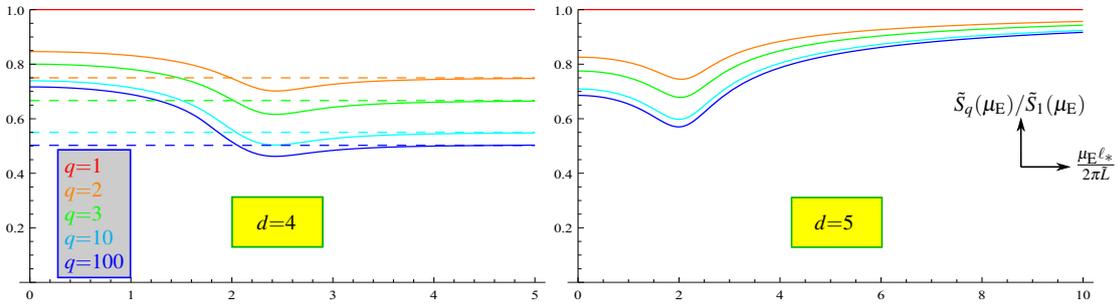
\begin{figure}[b]
\[
\raisebox{-58pt}{
  \begin{picture}(0,130)
  \put(-215,0){\scalebox{1.5}{\includegraphics{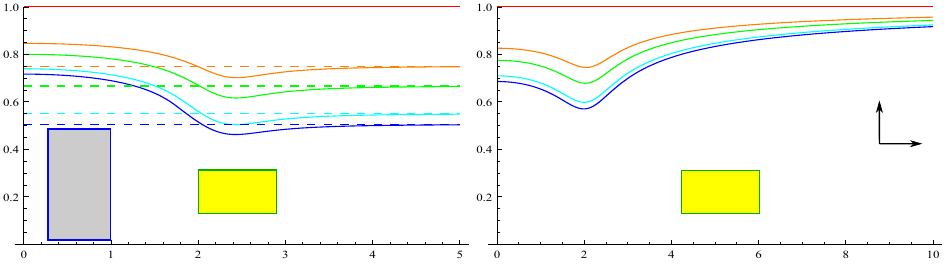}}}
  \put(0,5){
    \setlength{\unitlength}{.90pt}\put(-32,-1){
    \put(189,81) {$\scr \tilde{S}_q(\mu_\mt{E})/\tilde{S}_1(\mu_\mt{E})$}
    \put(239,56) {$\scr \frac{\mu_\mt{E}\ell_*}{2\pi \tilde{L}}$}
    \put(-181,55) {$\scr \color{red}q=1$}
    \put(-181,45) {$\scr \color{orange}q=2$}
    \put(-181,35) {$\scr \color{green}q=3$}
    \put(-181,25) {$\scr \color{cyan}q=10$}
    \put(-181,15) {$\scr \color{blue}q=100$}
    \put(-101,31) {$\scr d=4$}
    \put(131,31) {$\scr d=5$}
      }\setlength{\unitlength}{1pt}}
  \end{picture}}
\]
\caption{\small{\Renyi entropies normalized by $\tilde{S}_1 \left( \mu_\mt{E} \right)$ as a function of the imaginary chemical potential for the minimum allowed by causality Gauss-Bonnet coupling for $d=4$ and $d=5$. Each dashed line is for $S_q=\frac{q+1}{2q}$ with each $q$ given by the respective colour.}}\label{Fig:13}
\end{figure}

In figure \ref{Fig:13}, we plot ${{{{\tilde S}_q}\left( {{\mu_\mt{E}} } \right)}}/{{{{\tilde S}_1}\left( {{\mu} } \right)}}$ for the minimum allowed value of the Gauss-Bonnet coupling, versus the magnitude of the imaginary chemical potential for $d=4$ and $d=5$.
An interesting discrimination between the $d=4$ and $d>4$ cases can be observed, as long as the large imaginary chemical potential limit is concerned. This behaviour appears because black hole thermodynamics in Einstein-Gauss-Bonnet gravity are qualitatively different at $d=4$ in comparison to $d>4$, a fact that it is reflected in the zeroth order term of equation \eqref{eq:xn_equation} that is proportional to $d-4$. If one finds the large imaginary chemical potential limit of $x_q$ using equation \eqref{eq:xn_equation}, it can be shown that
\begin{equation}
\mathop {\lim }\limits_{{\mu_\mt{E}} \to \infty } \frac{{{{\tilde S}_q}\left( {{\mu_\mt{E}},\lambda } \right)}}{{{{\tilde S}_1}\left( {{\mu_\mt{E}},\lambda } \right)}} = \begin{cases} \cfrac{{q + 1}}{{2q}}, & d = 4 , \\
1, & d > 4 .
\end{cases}
\end{equation}

\subsection{\Renyi Entropy Inequalities}\label{subsec:ren entropy inequalities}

As discussed in subsection \ref{subsec:renyi_basics}, \Renyi entropies obey by definition a series of interesting inequalities. However, during the holographic calculation of \Renyi entropies, the probability distribution of their definition is not accessible, thus, the validity of these inequalities is not guaranteed. If the holographic bulk theory is meant to have a CFT dual at a stable thermal ensemble, the inequalities must still be obeyed. In the following, focusing to the inequality $\frac{\partial }{{\partial q}}\left( {\frac{{q - 1}}{q}{S_q \left( \mu \right)}} \right) \ge 0$ and following the analysis in \cite{Hung:2011nu}, we identify the consistency restrictions that have to be imposed in the bulk theory. Equation \eqref{eq:Renyi_replica_trick} implies that
\begin{equation}
\frac{\partial }{{\partial q}}\left( {\frac{{q - 1}}{q}{S_q \left( \mu \right)}} \right) = \frac{1}{{{q^2}}}S\left( {\frac{{{T_0}}}{q}} , \mu \right) .
\label{eq:renyi_second_inequality_holographic}
\end{equation}
As a result, the \Renyi entropy inequality under study holds only if the theory does not contain negative entropy black holes. Although the above is a trivial statement in Einstein gravity, in the presence of higher derivative corrections, it is not trivial due to corrections to the area law formula for the black hole entropy. It has been shown \cite{Anninos:2008sj} that in Gauss-Bonnet gravity, charged black holes with hyperbolic horizons are characterized by negative thermal entropy if the Gauss-Bonnet coupling and the chemical potential obey,
\begin{equation}
\lambda  > \frac{{\left( {d - 3} \right)\left( {{d^2} + d - 8} \right)}}{{4d{{\left( {d - 1} \right)}^2}}} \equiv \lambda_{cr},\quad
{{{\left( {\frac{{\mu {\ell _*}}}{{2\pi \tilde{L}}}} \right)}^2}} < \frac{1}{{{{\left( {d - 2} \right)}^2}}}\left( {\frac{{d{{\left( {d - 1} \right)}^2}}}{{d - 3}}\lambda  - \frac{{{d^2} + d - 8}}{4}} \right) .
\label{eq:lambda_negative_entropy_bound}
\end{equation}
\begin{wrapfigure}{r}{0.5\textwidth}
\vspace{-25pt}
\[
\raisebox{-58pt}{
  \begin{picture}(0,140)
  \put(-110,0){\scalebox{1.5}{\includegraphics{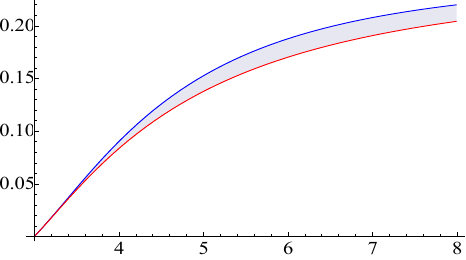}}}
  \put(0,5){
    \setlength{\unitlength}{.90pt}\put(-32,-1){
    \put(-78,123) {$\lambda$}
    \put(137,3) {$d$}
          }\setlength{\unitlength}{1pt}}
  \end{picture}}
\]
\vspace{-15pt}
\caption{\small{The bounds imposed on $\lambda$ by causality and by the positivity of the black hole entropy, as function of the number of dimensions.}}\label{fig:lambdabounds}
\vspace{-20pt}
\end{wrapfigure}
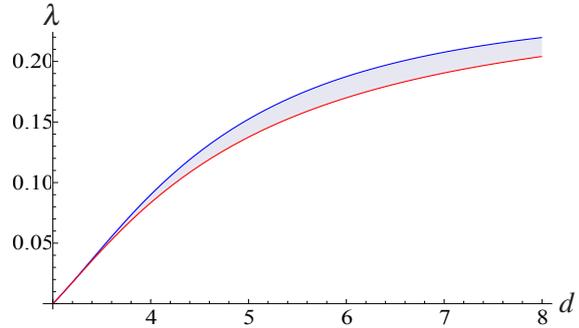
\noindent
The bound imposed to the Gauss-Bonnet coupling by \Renyi entropy inequalities is stricter than the bounds imposed by causality. This is depicted in figure \ref{fig:lambdabounds}, where couplings below the blue curve do not violate causality and couplings above the red curve allow for negative entropy black holes. The two bounds converge at $d=3$ and $d\to \infty$, but for any $d>3$ there is a range of Gauss-Bonnet couplings that do not violate causality and at the same time give rise to negative entropy black holes (shaded region).

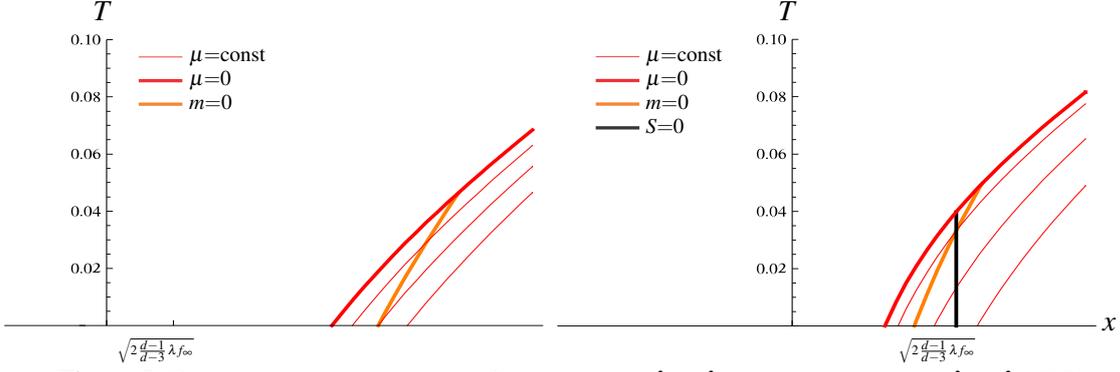
\begin{figure}[ht!]
\[
\raisebox{-58pt}{
  \begin{picture}(0,130)
  \put(-215,0){\scalebox{1.5}{\includegraphics{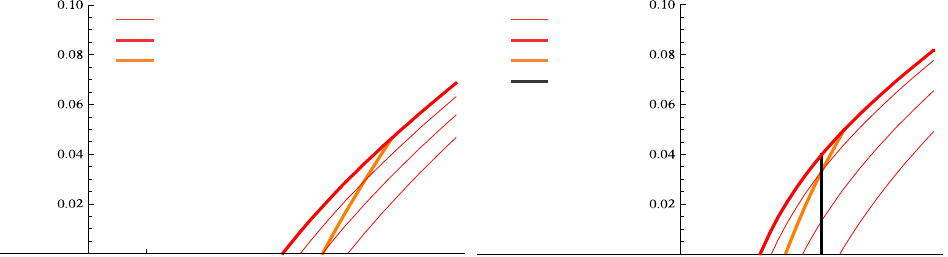}}}
  \put(0,5){
    \setlength{\unitlength}{.90pt}\put(-32,-1){
    \put(-170,125) {$T$}
    \put(-130,108) {$\scr\mu=\mathrm{const}$}
    \put(-130,98) {$\scr\mu=0$}
    \put(-130,88) {$\scr m=0$}
    \put(60,108) {$\scr\mu=\mathrm{const}$}
    \put(60,98) {$\scr\mu=0$}
    \put(60,88) {$\scr m=0$}
    \put(60,78) {$\scr S=0$}
    \put(115,125) {$T$}
    \put(250,-5) {$x$}
    \put(-160,-15) {\tiny{$\scr\sqrt{2 \frac{d-1}{d-3} \lambda f_\infty}$}}
    \put(165,-15) {\tiny{$\scr\sqrt{2 \frac{d-1}{d-3} \lambda f_\infty}$}}
          }\setlength{\unitlength}{1pt}}
  \end{picture}}
\]
\caption{\small{Black hole thermodynamics. On the l.h.s for $\lambda  < \lambda_{cr}$ and on the r.h.s for $\lambda  > \lambda_{cr}$ \cite{Anninos:2008sj}.}}\label{fig:bhthermodynamics}
\end{figure}

Equation \eqref{eq:black_hole_thermal_entropy} suggests that black holes characterized by negative thermal entropy must obey (figure \ref{fig:bhthermodynamics}),
\begin{equation}
{x^2} < x_{S = 0}^2 \equiv 2\frac{{d - 1}}{{d - 3}}\lambda {f_\infty } .
\end{equation}
As one can observe in figure \ref{fig:bhthermodynamics}, all black holes with negative entropy have smaller Hawking temperature than the temperature $T\left( {{x_{S = 0}},\mu } \right)$ of the black hole with vanishing entropy.

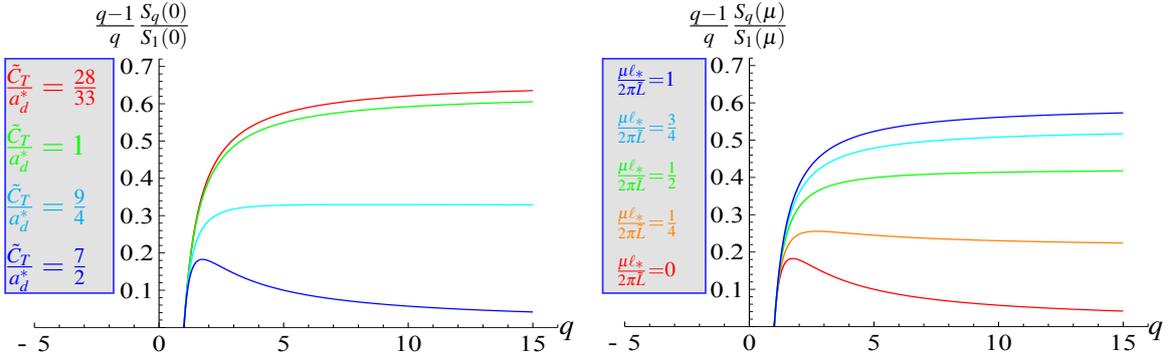
\begin{figure}[ht!]
\[
\raisebox{-58pt}{
  \begin{picture}(0,130)
  \put(-215,0){\scalebox{1.5}{\includegraphics{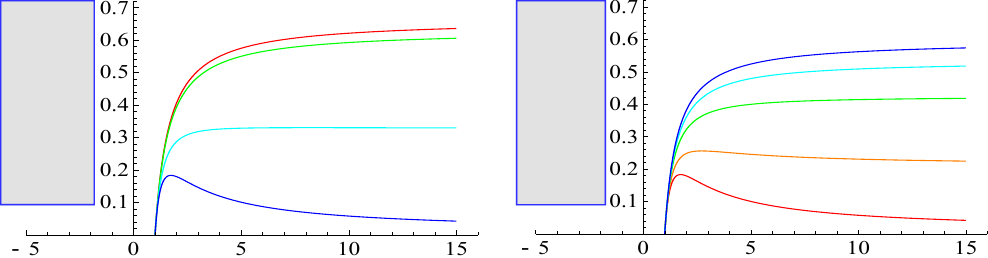}}}
  \put(0,5){
    \setlength{\unitlength}{.90pt}\put(-32,-1){
    \put(-170,130){$\frac{q-1}{q}\frac{S_q(0)}{S_1(0)}$}
    \put(77,130)  {$\frac{q-1}{q}\frac{S_q(\mu)}{S_1(\mu)}$}
    \put(269,4)   {$q$}
    \put(24,4)    {$q$}
    \put(-207,104){\color{red}$\frac{\tilde{C}_T}{a_d^*}=\frac{28}{33}$}
    \put(-207,79) {\color{green}$\frac{\tilde{C}_T}{a_d^*}=1$}
    \put(-207,54) {\color{cyan}$\frac{\tilde{C}_T}{a_d^*}=\frac{9}{4}$}
    \put(-207,29) {\color{blue}$\frac{\tilde{C}_T}{a_d^*}=\frac{7}{2}$}
    \put(47,108)  {\color{blue}$\scr \frac{\mu\ell_*}{2\pi \tilde{L}}=1$}
    \put(47,88)   {\color{cyan}$\scr \frac{\mu\ell_*}{2\pi \tilde{L}}=\frac{3}{4}$}
    \put(47,68)   {\color{green}$\scr \frac{\mu\ell_*}{2\pi \tilde{L}}=\frac{1}{2}$}
    \put(47,48)   {\color{orange}$\scr \frac{\mu\ell_*}{2\pi \tilde{L}}=\frac{1}{4}$}
    \put(47,28)   {\color{red}$\scr \frac{\mu\ell_*}{2\pi \tilde{L}}=0$}
          }\setlength{\unitlength}{1pt}}
  \end{picture}}
\]
\caption{\small{The violation of second inequality obeyed by \Renyi entropies for $d=7$.}}\label{fig:inequality_violation}
\end{figure}

According to the above, it is inevitable that for high enough $q$, ${T_0}/{q}$ will become eventually smaller than the above critical value, corresponding to a negative entropy black hole and giving rise to a violation of the second inequality for \Renyi entropies as described by equation \eqref{eq:renyi_second_inequality_holographic}. More specifically this occurs for $q > q_v$, where
\begin{equation}
{q_v} = \frac{{{T_0}}}{{T\left( {{x_{S = 0}},\mu } \right)}} = \frac{{8\sqrt {2\frac{{d - 1}}{{d - 3}}\lambda {f_\infty }} }}{{\frac{{4d{{\left( {d - 1} \right)}^2}}}{{d - 3}}\lambda  - \left( {{d^2} + d - 8} \right) - {{\left( {d - 2} \right)}^2}{{\left( {\frac{{\mu {\ell _*}}}{{2\pi \tilde{L}}}} \right)}^2}}}
\end{equation}
This is depicted in figure \ref{fig:inequality_violation}, where $\frac{q-1}{q}\frac{S_q}{S_1}$ is plotted versus $q$ for vanishing chemical potential and for several Gauss-Bonnet couplings covering all the region allowed by causality on the l.h.s. and for the maximum value of Gauss-Bonnet coupling allowed by causality and several chemical potentials on the r.h.s.

\section{Discussion}
\label{sec:discussion}

We managed to perform a holographic calculation of entanglement \Renyi entropies for spherical entangling surfaces in CFT's with a conserved charge and Einstein-Gauss-Bonnet holographic duals, extending the results of \cite{Belin:2013uta} to theories that do not necessarily obey $c=a$. The entanglement \Renyi entropy is calculated using appropriate conformal transformations that link the latter with the thermal entropy in the hyperbolic cylinder as introduced in \cite{Casini:2011kv} and applied in theories without conserved charges in \cite{Hung:2011nu}.

Entanglement \Renyi entropies depend approximately linearly on the central charge ratio ${\tilde{C}_T}/{a_d^*}$ in the regime consistent with causality, similarly to the findings of \cite{Hung:2011nu}. However, unlike the uncharged case, it is not always true that they acquire their maximum value for the minimum value of ${\tilde{C}_T}/{a_d^*}$. Above a finite critical chemical potential, which is an increasing function of $q$, the monotonicity of the aforementioned approximately linear dependence of $S_q$ on ${\tilde{C}_T}/{a_d^*}$ gets inverted and as a result the maximum value for $S_q$ is achieved for the maximum value of the ratio ${\tilde{C}_T}/{a_d^*}$.

Entanglement \Renyi entropies are an increasing function of the chemical potential. For large chemical potentials, entanglement \Renyi entropy becomes independent of $q$, in agreement with the results of \cite{Belin:2013uta} and it is completely determined by the product $\sqrt{f_\infty} \mu$. The dependence of entanglement \Renyi entropy on an imaginary chemical potential indicates the existence of some kind of phase transition as it is qualitatively different for $\lambda \geq 0$ and $\lambda<0$. Specifically, the analytic continuation of entanglement \Renyi entropy is well defined for arbitrarily high imaginary chemical potential when $\lambda<0$; this is not the case for $\lambda \geq 0$, where at a finite imaginary chemical potential a branch cut appears. The latter is related to the topological black hole in the bulk theory turning to a naked singularity at this specific imaginary chemical potential.

A very interesting outcome of the holographic calculation of entanglement \Renyi entropies is the violation of an inequality they must obey by definition, above a critical value of the Gauss-Bonnet coupling, $\lambda_{cr}  = \frac{{( {d - 3} )( {{d^2} + d - 8} )}}{{4d{{( {d - 1} )}^2}}}$. The validity of the aforementioned inequality is holographically related to the positivity of the entropy of the related topological black holes. The latter is not proportional to the horizon area in the presence of higher derivative corrections, but it is calculated on the basis of Wald's formula, which may return a negative entropy for $\lambda > \lambda_{cr}$. Although $\lambda_{cr}$ is very close to the maximum $\lambda$ allowed by causality, it is smaller than that for any $d>3$, allowing a range of $\lambda$ that are consistent with causality and give rise to \Renyi entropy inequality violation. Consequently, gravitational theories with higher derivative corrections, holographically connected to a CFT at a stable thermal ensemble should be restricted by a constraint stricter than causality.

We have to point out that the existence of negative entropy black holes and the consequent violation of \Renyi entropy inequality are both an artifact of higher derivative terms competing with the leading Einstein term. The proximity of the Gauss-Bonnet coupling bound by \Renyi entropy inequality to the bound by causality suggests that in a realistic M-theory or string theory compactification, where even higher order terms have to be taken into account, causality bound can easily become stricter (or equally strict if there is an underlying reason) to the \Renyi inequality bound, thus not allowing \Renyi inequality violations.

Another possible resolution to the existence of both negative entropy black holes and \Renyi entropy inequality violation may be provided in \cite{Cvetic:2001bk}. The authors conjecture that since the parameters that allow asymptotically AdS black holes with negative entropy correspond to asymptotically dS black holes with positive entropy, some kind of phase transition between the asymptotically AdS and asymptotically dS black hole occurs. If this conjecture is true, then this region of parameters may contain information about the holographic description of dS space.

An interesting extension of the present work would be the inclusion of higher derivative gauge interactions. Black hole thermodynamics in the presence of such interactions have been studied in \cite{Anninos:2008sj}. It may also be interesting to study entanglement \Renyi entropies in the canonical ensemble, where the charge instead of the chemical potential is kept fixed.

\acknowledgments

This work is supported and implemented under the ARISTEIA action of the operational programme for education and long life learning and is co-funded by the European Union (European Social Fund) and National Resources of Greece.

\end{document}